\documentclass[aps,prc,twocolumn,superscriptaddress,showpacs,amsmath,floatfix]{revtex4}

\usepackage{graphicx}
\usepackage{bm}
\usepackage{CJKutf8}
\newcommand{\ba}{\begin{array}}
\newcommand{\ea}{\end{array}}
\newcommand{\rmd}{{\rm d}}

\newcommand{\Tr}{{\rm Tr}}

\begin{document}
%\begin{CJK*}{GBK}{song}
\begin{CJK*}{GB}{gbsn}

%\title{Particle number restoration in energy-density-functional approach by Lipkin method}
%\title{Lipkin method of particle-number restoration to high orders}
\title{Lipkin method of particle-number restoration to higher orders}

\author{X.B. Wang (ÍõС±£)}
\affiliation{Department of Physics, P.O. Box 35 (YFL), University of Jyv$\ddot{a}$skyl$\ddot{a}$, FI-40014 Jyv$\ddot{a}$skyl$\ddot{a}$, Finland}
\author{J. Dobaczewski}
\affiliation{Department of Physics, P.O. Box 35 (YFL), University of Jyv$\ddot{a}$skyl$\ddot{a}$, FI-40014 Jyv$\ddot{a}$skyl$\ddot{a}$, Finland}
\affiliation{Helsinki Institute of Physics, P.O. Box 64, FI-00014 University of Helsinki, Finland}
\author{M. Kortelainen}
\affiliation{Department of Physics, P.O. Box 35 (YFL), University of Jyv$\ddot{a}$skyl$\ddot{a}$, FI-40014 Jyv$\ddot{a}$skyl$\ddot{a}$, Finland}
\affiliation{Helsinki Institute of Physics, P.O. Box 64, FI-00014 University of Helsinki, Finland}
\author{L.F. Yu (ÓàÁéåú)}
\affiliation{Department of Physics, P.O. Box 35 (YFL), University of Jyv$\ddot{a}$skyl$\ddot{a}$, FI-40014 Jyv$\ddot{a}$skyl$\ddot{a}$, Finland}
\author{M.V. Stoitsov}
\affiliation{Department of Physics \& Astronomy, University of Tennessee, Knoxville, Tennessee 37996, USA and \\
Physics Division, Oak Ridge National Laboratory, Oak Ridge, Tennessee 37831, USA}

\begin{abstract}

\begin{description}
\item[Background:]
On the mean-field level, pairing correlations are incorporated
through the Bogoliubov-Valatin transformation, whereupon the particle
degrees of freedom are replaced by quasiparticles.
This approach leads to a spontaneous breaking of the particle-number
symmetry and mixing of states with different particle numbers. In
order to restore the particle number, various methods have been
employed, which are based on projection approaches before or after
variation. Approximate variation-after-projection (VAP) schemes,
utilizing the Lipkin method, have mostly been used within the
Lipkin-Nogami prescription.

\item[Purpose:]
Without recurring to the Lipkin-Nogami prescription, and using
instead states rotated in the gauge space, we derive the Lipkin
method of particle-number restoration up to sixth order and we test
the convergence and accuracy of the obtained expansion.

\item[Methods:]
We perform self-consistent calculations using the higher-order Lipkin
method to restore the particle-number symmetry in the framework of
superfluid nuclear energy-density functional theory. We also apply
the Lipkin method to a schematic exactly solvable two-level pairing
model.

\item[Results:]
Calculations performed in open-shell tin and lead isotopes show that
the Lipkin method converges at fourth order and satisfactorily
reproduces the VAP ground-state energies and energy kernels. Near
closed shells, the higher-order Lipkin method cannot be applied because of
a non-analytic kink in the ground-state energies in function of the
particle number.

\item[Conclusions:]
In open-shell nuclei, the higher-order Lipkin method provides a good
approximation to the exact VAP energies. The method is computationally
inexpensive, making it particularly suitable, for example, for future
optimizations of the nuclear energy-density functionals and
simultaneous restoration of different symmetries.

\end{description}

\end{abstract}

\pacs{21.10.-k,21.60.Jz,74.20.-z,71.15.Mb}

\date{\today}
\maketitle
\end{CJK*}

\section{Introduction}

Ground states of the atomic nuclei exhibit nucleonic pairing
correlations. This manifests itself in odd-even mass staggering,
properties of low-lying excited states, moments of inertia, etc., to name
but a few examples~\cite{[Boh69w],(Rin00)}.
To successfully describe these phenomena, nucleonic pairing
is usually introduced within mean-field models and handled through the
Bogoliubov-Valatin transformation~\cite{(Rin00)}, whereupon the particle degrees of freedom
are replaced by quasiparticles.
This effectively incorporates the pairing correlations but, as a consequence,
leads to particle-number-mixed wave functions.
Situation is the same also for other symmetries broken on the mean-field level,
e.g., by allowing nucleus to deform, quadrupole-type correlations are effectively
incorporated in the mead-field picture, at the expense of breaking the rotational invariance.
Nevertheless, true ground states conserve all symmetries of the underlying
Hamiltonian, including the particle number.

To link the spontaneous breaking of symmetries to symmetry-conserving states,
various symmetry-restoration schemes have been utilized.
In principle, in self-consistent approaches solved within iterative methods,
broken symmetries should
be restored during every step towards solution. This is the well know
variation-after-projection (VAP)~\cite{[Ang01],[Sto07],[Hup11]} method.
The drawback of VAP is computationally expensive integration over the gauge angles
of symmetries to be restored, applied during every iteration step.
Therefore, usual practice is to revert to computationally less intensive
projection-after-variation (PAV)~\cite{(Rin00),[Ben03]} scheme, where
symmetries are restored at the end, from the converged self-consistent symmetry-broken
mean-field solution.

When applying VAP method to superfluid nuclear density functional theory (DFT), some of the
energy density functionals (EDFs) seem to be ill-suited for the task. In particularly,
with widely used Skyrme-like EDFs, several pathologies exists. With particle
number restoration, poles and non-analytic behavior preclude obtaining a unique
solution~\cite{[Dob07d],[Lac09],[Ben09]}. These difficulties are traced to the non-integer powers
of density in the employed EDFs~\cite{[Dob07d],[Dug09]}. The same also holds for the
restoration of angular momentum~\cite{[Sat11b]}.

To circumvent prohibitive computational cost of the VAP method, an
approximate method is called for. The central rationale in the Lipkin
method, which fulfill this goal, is to replace the original Hamiltonian by an auxiliary
Routhian, making the symmetry-projected states degenerate in
energy~\cite{[Lip60]}. This allows us to approximately evaluate
ground-state properties of the corresponding symmetry-restored system
without actually performing any projection~\cite{[Lip60],[Dob09g]}.
In particularly for the particle number restoration, a power series
expansion as a function of particle number fluctuation was suggested.
Based on this idea, Nogami~\cite{[Nog64],[Pra73]} introduced a
prescription to calculate coefficients of the power expansion at
second order, which is called the Lipkin-Nogami (LN) method, and which has
been widely used in nuclear DFT calculations~\cite{[Ben03],[Flo97]}.

Quantitative effect of the particle-number-restoration
largely depends on whether the pairing correlations are strong
(mid-shell) or weak (near closed shell). As pointed out in
Refs.~\cite{[Zhe92],[Dob93],[Val00],[She02]}, the parabolic approximation,
which corresponds to a sum up to the second order of the Lipkin or
Kamlah~\cite{[Kam68]} approximation, may fail at the limit of weak pairing.
In this work we extend the Lipkin method beyond second order used
so far, so as to make the first tests of its convergence and accuracy.

The central issue of the Lipkin method is to search for a suitable set
of the Lipkin power-expansion parameters~\cite{[Dob09g]}. For the LN method, the
second-order parameter is calculated through the diagonal matrix
elements~\cite{[Nog64],[Pra73],[Rei96],[Ben00e]}, which requires us to calculate the
linear response of the mean field to the particle-number projection~\cite{[Rei96],[Ben00e]}.
However, the response term, which has large influence on potential
energy surface (PES)~\cite{[Rei96]}, is cumbersome to evaluate. Therefore,
usually in calculations involving the LN method, an approximate prescription of
seniority pairing is used to obtain the effective pairing strength for
the second-order term~\cite{[Sto05],[Sto07]}.
In this work, we propose a different way to derive these
expansion parameters, namely, starting from the non-diagonal energy kernels.

This paper is organized as follows: In Sec.~\ref{sec2}, we cover our
theoretical framework of the Lipkin method for particle number
projection. In Sec.~\ref{sec3}, we present numerical results, and in Sec.~\ref{sec4}, we
give the summary and outlook. Appendix~\ref{app1}
contains explicit expressions of the Lipkin method applied in this
work and Appendix~\ref{app2} provides an illustration of the method
within the exactly solvable two-level pairing model.

\section{Lipkin method}
\label{sec2}

To start, we first recall some of the standard definitions available
in the literature, which are required in the present work. Within the
Hartree-Fock-Bogoliubov (HFB) framework, the wave function rotated in
the gauge space is defined as~\cite{(Rin00),[Ben03]}
\begin{equation}
\label{eq:phi}
|\Phi(\phi)\rangle =
\exp\left(i\phi(\hat{N}-N_0)\right)|\Phi\rangle \, ,
\end{equation}
where $\phi$ is the gauge angle, $\hat{N}$ is the particle-number operator,
and $N_0=\langle\Phi|\hat{N}|\Phi\rangle$ is the average particle number.
In what follows, for a sake of clarity, we present expressions for a system
composed only of one kind of particles. Generalizations to two types of particles,
that is, to protons and neutrons, is straightforward and is discussed briefly later.
Similarly as in Ref.~\cite{[Dob09g]}, the overlap and energy kernels are defined as
\begin{eqnarray}
\label{eq:i}
I(\phi) &=& \langle\Phi|\Phi(\phi)\rangle \, , \\
\label{eq:h}
H(\phi) &=& \langle\Phi|\hat{H}|\Phi(\phi)\rangle \, ,
\end{eqnarray}
and kernels of $(\hat{N}-N_0)^m$ as
\begin{equation}
\label{eq:nm}
N_m(\phi) =\langle\Phi|(\hat{N}-N_0)^m|\Phi(\phi)\rangle \, .
\end{equation}
The kernels of Eq. (\ref{eq:nm}) can be calculated as
derivatives of the overlap kernel with respect to the gauge angle
\begin{equation}
N_m(\phi) =  (-i)^m\frac{d^m}{d\phi^m}I(\phi) \, .
\label{eq:Nm}
\end{equation}
Explicit expression for these kernels are presented in Appendix \ref{app1}.
In Eqs.~(\ref{eq:h}) and (\ref{eq:nm}), kernels are defined
in terms of matrix elements. However, within the EDF methods they
have to be understood as standard functions of transition density
matrices, see, e.g., discussion in Ref.~\cite{[Dob07d]}.

As demonstrated by Lipkin~\cite{[Lip60]}, variation after the particle-number projection
(VAPNP) can be performed using an auxiliary Routhian,
\begin{equation}
\label{eq:calK}
\hat{H}'
= \hat{H}-\hat{K}\{\hat{N}-N_0\} \, ,
\end{equation}
where Lipkin operator $\hat{K}$, which is a function of the
shifted particle-number operator $\hat{N}-N_0$, is chosen so as to
``flatten'' the $N$-dependence of average Routhians calculated for the particle-number
projected states~\cite{[Lip60],[Dob09g]}. Had these projected
Routhians been exactly $N$-independent (perfectly flat), the exact projected energy
$E_{N_0}$ could have been obtained by minimizing the average value
of the Routhian for the unprojected state $|\Phi\rangle$, that is,
\begin{equation}
\label{eq:K-eig}
E_{N_0}=\langle\Phi|\hat{H}-\hat{K}\{\hat{N}-N_0\}|\Phi\rangle \, .
\end{equation}
Otherwise, the Lipkin method constitutes a suitable approximate VAPNP
method, and its accuracy depends on the quality of the choice made
for the Lipkin operator $\hat{K}$.

As suggested by Lipkin~\cite{[Lip60]}, the simplest and manageable
ansatz for the Lipkin operator $\hat{K}$ has the form of a power
expansion,
\begin{equation}
\label{eq:fN}
\hat{K}\{\hat{N}-N_0\}=  \sum_{m=1}^M  k_{m} (\hat{N}-N_0)^m \, ,
\end{equation}
where $k_{1}\equiv\lambda$ is the Fermi energy, which is used to fix the
average particle number. Furthermore, higher-order Lipkin parameters $k_{m}$ are used to
best describe the particle-number dependence of the average energies of
projected states.

Up to now, the LN method %\cite{[Nog64],[Pra73],[Rei96],[Ben00e]}
was frequently used to estimate values of $k_{2}$ (traditionally denoted by $\lambda_{2}$).
However, this method relies on calculating the average values of
$\langle\Phi|{\hat H} \hat{N}^m|\Phi\rangle$ and
$\langle\Phi|\hat{N}^m|\Phi\rangle$, and, thus, at higher
orders ($m>2$) evaluation of these terms becomes cumbersome and
impractical.

The essence of the original Lipkin method is different, namely, it relies
on deriving expressions
for $k_{m}$ that ``flatten'' the $\phi$-dependence
of the reduced Routhian kernel $h'(\phi)$, that is,
\begin{equation}
\label{flat}
h'(\phi) = h(\phi) - \sum_{m=1}^M  k_{m} n_m(\phi) \, ,
\end{equation}
where
\begin{equation}
\label{reduced-kernels}
h' (\phi)=\frac{H' (\phi)}{I(\phi)} \, , ~
h  (\phi)=\frac{H  (\phi)}{I(\phi)} \, , ~
n_m(\phi)=\frac{N_m(\phi)}{I(\phi)} \, .
\end{equation}
%where $h'(\phi)=H'(\phi)/I(\phi)$,  $h(\phi)=H(\phi)/I(\phi)$, and $n_m(\phi)=N_m(\phi)/I(\phi)$.
Up to any order, this is a perfectly manageable setup, because for an arbitrary
value of the gauge angle, the generalized Wick theorem~\cite{(Rin00)}
allows for a simple determination of the energy and overlap kernels $H(\phi)$
and $N_m(\phi)$.
Explicit expressions for $k_{m}$ are presented in Appendix~\ref{app1}.

For a perfectly flat ($\phi$-independent) reduced Routhian kernel $h'(\phi)\equiv{C}$,
we then have the exact average value of the Routhian evaluated for the state
projected on particle number $N_0$,
\begin{eqnarray}
\label{eq:PNP}
E'_{N_0} &=& \frac{\int_0^{2\pi} H'(\phi)\rmd\phi}
                  {\int_0^{2\pi}  I(\phi)\rmd\phi}% \nonumber \\
          =  \frac{\int_0^{2\pi} h'(\phi)I(\phi)\rmd\phi }
                  {\int_0^{2\pi}  I(\phi)\rmd\phi} = C \, .
\end{eqnarray}
Since for the state projected on $N_0$, the average value of the Lipkin
operator (\ref{eq:fN}) is, by definition, equal to zero, we also have that
\begin{eqnarray}
E_{N_0} &=&  C,
\end{eqnarray}
and thus the minimization of the average Routhian (\ref{eq:K-eig}) is
equivalent to the exact VAPNP. Again, any imperfection in the
$\phi$-independence of $h'(\phi)$ amounts to a certain approximation
of the exact VAPNP. However, since it is now relatively easy to go to
higher orders in the power expansion of Eq.~(\ref{eq:fN}), we can
systematically test the convergence of this expansion.

The largest contributions to integrals
in Eq.~(\ref{eq:PNP}) come
from the vicinity of the origin due to the largest
weight~\cite{[Flo97]}. Therefore, we can evaluate Lipkin parameters $k_{m}$
using the gauge-rotated intrinsic states near the origin. This also
avoids the singularities caused by vanishing overlaps~\cite{[Dob07d]}. As
an example, at second order one obtains the Lipkin parameter,
\begin{equation}
\label{eq:lambda2-lipkin}
k_{2}=\frac{h(\phi_2)-k_{1}n_1(\phi_2)-h(0)}{n_2(\phi_2)-n_2(0)} \, ,
\end{equation}
where $\phi_2$ is a pre-selected small value of the gauge angle,
and the flattened energy reads
\begin{equation}
\label{eq:hlambda2}
E_{N_0} =\frac{h(0)n_2(\phi_2)-h(\phi_2)n_2(0)+k_{1}n_1(\phi_2)n_2(0)}{n_2(\phi_2)-n_2(0)} \, .
\end{equation}
Had the expansion up to second order been exact, values of $k_{2}$
and $E_{N_0}$ obtained from Eqs.~(\ref{eq:lambda2-lipkin}) and
(\ref{eq:hlambda2}) would have been independent of $\phi_2$. Thus,
their eventual dependence on $\phi_2$ indicates the necessity of going
beyond second order.

Similarly, at order $M$, we evaluate Lipkin parameters $k_{m}$,
$m=1,\ldots,M$, using a set of $M$ small gauge angles $\phi_i$,
$i=1,\ldots,M$. In practice, in this work, we use equally spaced
values of $\phi_i=i\phi_1$, and at each order we check the eventual
dependence of results on the maximum gauge angle used, $\phi_{M}$.

We note here that the minimization of the average Routhian
(\ref{eq:K-eig}) with respect to the HFB state $|\Phi\rangle$ can be
performed by solving the standard HFB equation with additional
higher-order terms added, see Appendix~\ref{app1}. We also note that
Lipkin parameters $k_{m}$ must be determined in each HFB iteration
(for each current state $|\Phi\rangle$), in such a way that at the end
of the HFB convergence they correspond to the final self-consistent
solution, and thus parametrically depend on it. However, this
dependence does not give rise to any additional terms in the HFB equation,
because the derivation of the Lipkin method is based on treating them as constants,
cf.\ discussion of the LN and Kamlah methods in Ref.~\cite{[Dob93]}.

An exactly solvable two-level pairing model offers an ideal
environment to test qualitative properties of the Lipkin VAPNP
method. The results presented in Appendix~\ref{app2} show that in
such a schematic model, the higher-order Lipkin VAPNP method is able
to reproduce correctly the exact VAPNP ground-state energies, both in
weak and strong pairing regimes, everywhere apart from the immediate
vicinity of the closed shell. This gives us confidence in applications
of this method in more involved cases of actual nuclei, which is
discussed in the next section.

\section{Results and discussion}
\label{sec3}

We have implemented the Lipkin VAPNP method, presented in Sec.~II, in
computer code {\sc HFODD} (v2.68c)~\cite{[Sch12],[Sch14]}. This code
solves the HFB equations in a three-dimensional Cartesian harmonic oscillator basis.
Within this implementation, we tested the Lipkin VAPNP method using the Skyrme
SIII parameterization~\cite{[Bei75]} in the particle-hole channel and
volume zero-range pairing interaction in the particle-particle channel.
SIII parameterization was selected for this study due to the fact that
it contains only integer powers of densities.
Used neutron pairing strength, $V_0=-155.45$\,MeV\,fm$^3$, was adjusted
within the LN method to reproduce the empirical neutron pairing gap of
$\Delta_n=1.245$ MeV of $^{120}$Sn.

In principle, at each given order of the Lipkin VAPNP method, this
adjustment should be repeated. However, for the sake of meaningful comparison
of the results obtained at different orders, we use the same pairing strength throughout all calculations.

For protons, pairing strength was set to zero, that is, the proton
subsystem is described by unpaired states. This setup allows us to
test the Lipkin VAPNP method in the neutron paired subsystem,
resulting in a clearer interpretation of the obtained results and
allowing for a better evaluation of the efficiency of the Lipkin
VAPNP method. Because of the used zero-range pairing interaction, we
adopted the commonly used equivalent-spectrum cutoff of 60\,MeV,
applied in the quasiparticle configuration space. All calculations
were performed in the spherical basis of 14 major harmonic-oscillator
shells.

\begin{figure}
\includegraphics[angle=0,width=0.9\columnwidth]{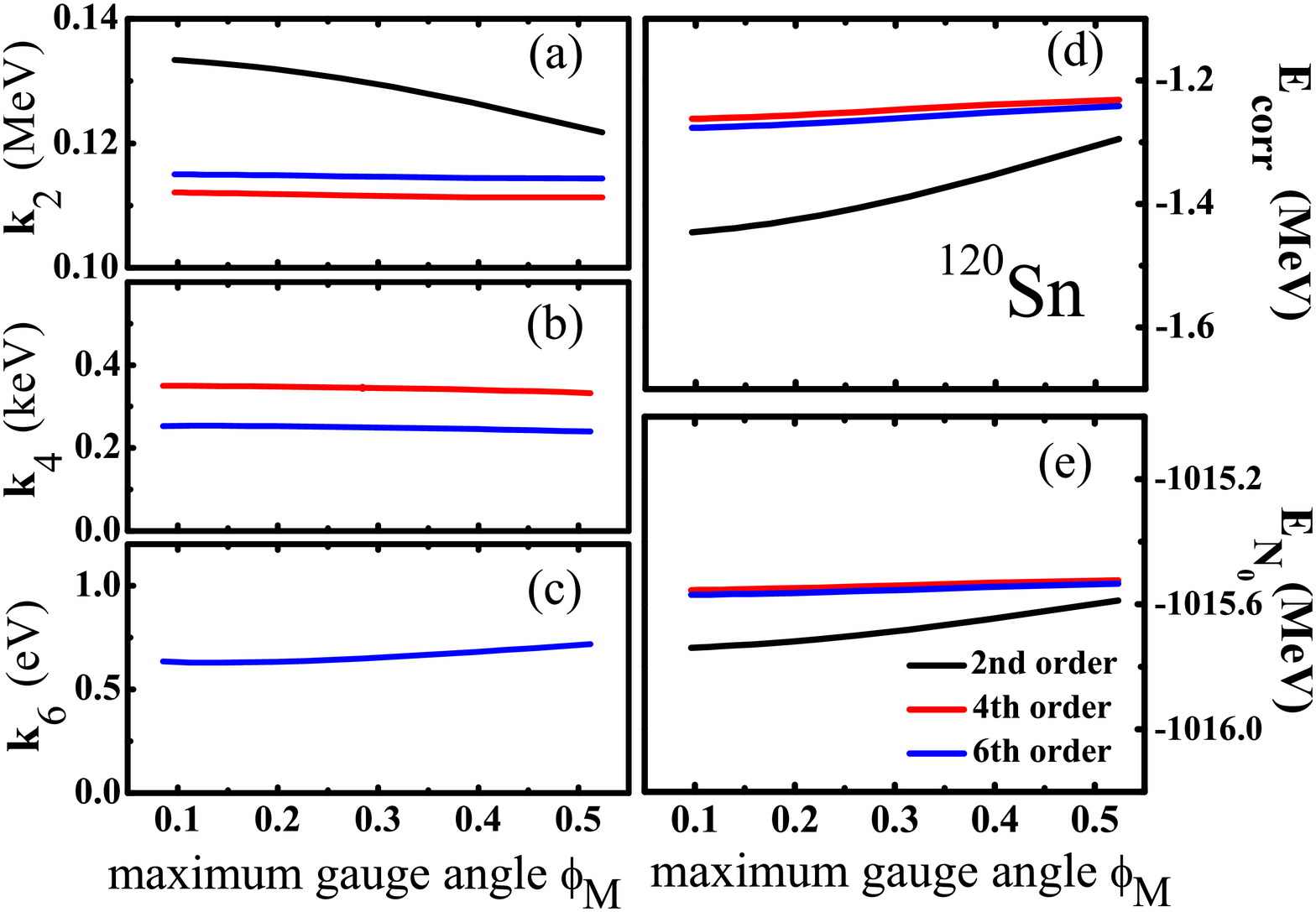}
\caption[T]{\label{fig:convergence}
(Color online) Lipkin parameters $k_2$ (a), $k_4$ (b), and $k_6$ (c),
Lipkin correction energy $E_{\text{corr}}$ (\protect\ref{eq:correction}) (d),
and Lipkin VAPNP energy $E_{N_0}$ (\protect\ref{eq:K-eig}) (e),
determined in $^{120}$Sn at second, fourth, and
sixth orders, as functions of the maximum gauge angle $\phi_{M}$.
Note that Lipkin parameters $k_2$, $k_4$, and $k_6$ are shown in units
of MeV, keV, and eV, respectively, which illustrates the rapid convergence
of the Lipkin expansion.
}
\end{figure}

\begin{figure}
\includegraphics[angle=0,width=0.9\columnwidth]{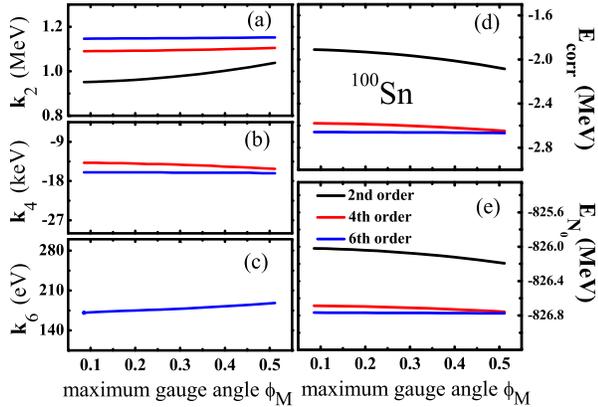}
\caption[T]{\label{fig:convergence-sn100}
(Color online) Same as Fig.~\ref{fig:convergence}, but for $^{100}$Sn.}
\end{figure}

To begin with, we
first study convergence of the Lipkin VAPNP method when terms up to
sixth-order in expansion (\ref{eq:fN}) are incorporated. At present, we limit our analysis to the even
powers only, that is, we take into account terms with $m=2$, 4, and 6.
This corresponds to a symmetric approximation around the central
value of the particle number $N_0$. In Figs.~\ref{fig:convergence}
and~\ref{fig:convergence-sn100}, we show, respectively for $^{120}$Sn
and $^{100}$Sn, dependence of the Lipkin parameters on the maximum gauge
angle $\phi_{M}$ (see the previous section). The figures also show the
total Lipkin VAPNP energy $E_{N_0}$ and
Lipkin correction energy $E_{\text{corr}}$
%to $\langle\Phi|\hat{H}|\Phi\rangle$,
\begin{equation}
\label{eq:correction}
E_{\text{corr}}=\langle\Phi|-\hat{K}\{\hat{N}-N_0\}|\Phi\rangle
= -\sum_{m=1}^M  k_{m} n_m(0) \, ,
\end{equation}
cf.~Eqs.~(\ref{eq:K-eig}) and~(\ref{flat}).

At second order, the obtained results show a clear dependence on $\phi_{M}$, indicating
insufficient expansion. On the other hand, at fourth and sixth orders,
total energy $E_{N_0}$ and correction energy $E_{\text{corr}}$
are already rather insensitive to $\phi_{M}$. Thus, we can conclude that at sixth order,
the expansion is well converged, and at least fourth order
is required for sufficiently precise results.
We also note that for the magic nucleus $^{100}$Sn, the convergence
is slightly slower, and the values of Lipkin parameters
are significantly higher than those for $^{120}$Sn.

\begin{figure}
\includegraphics[angle=0,width=0.9\columnwidth]{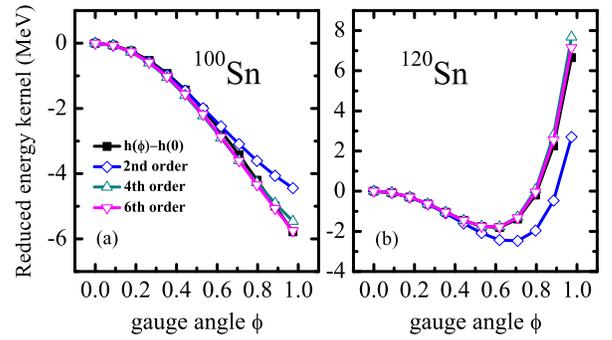}
\caption[T]{\label{fig:hphi-s}(Color online) Reduced energy kernel
$h(\phi)-h(0)$ (full squares) and reduced kernels of Lipkin
operator $\sum_{m=1}^M k_m(n_m(\phi)-n_m(0))$ at orders $M=2$, 4, and
6 (open symbols), as functions of the gauge angle up to $\phi=1$,
calculated in $^{100}$Sn and $^{120}$Sn.
}
\end{figure}

\begin{figure}
\includegraphics[angle=0,width=0.9\columnwidth]{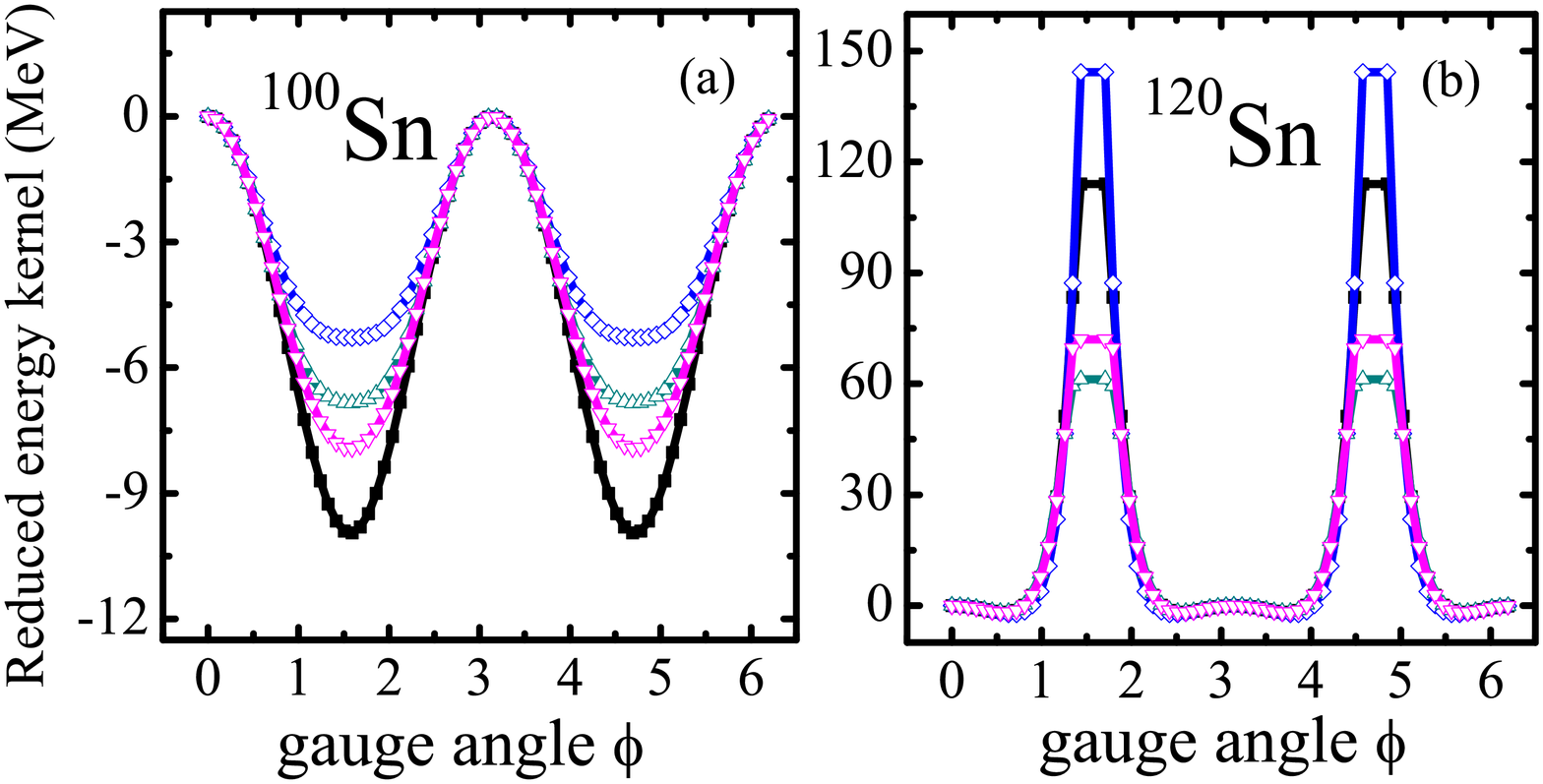}
\caption[T]{\label{fig:hphi-2}
(Color online) Same as in Fig.~\protect\ref{fig:hphi-s} but for
the gauge angles up to $\phi=2\pi$.
}
\end{figure}

In what follows, we have used the same maximum gauge angle of
$\phi_{M}=\frac{2\pi}{51}\simeq0.123$ in all expansions, regardless
of the expansion order.
In Fig.~\ref{fig:hphi-s}, convergence of the reduced kernels of
Lipkin operator (\ref{eq:fN}) in $^{100}$Sn and $^{120}$Sn is shown.
Kernel values at $\phi=0$ were subtracted, in order to illustrate how
well the reduced Routhian kernels (\ref{flat}) stay constant, that is,
independent of the gauge angle $\phi$.
Again, we clearly see that the second-order expansion is insufficient,
whereas fourth and sixth orders already give satisfactory description of the energy
kernels.

Figure~\ref{fig:hphi-2} shows the same kernels as those plotted in
Fig.~\ref{fig:hphi-s} for the whole range of gauge angle,
up to $\phi=2\pi$.
We see that in $^{100}$Sn the energy
kernel near $\phi=\pi/2$ is poorly described by the Lipkin expansion.
This is directly related to the kink in the particle-number dependence of the
projected energies, which appears at the magic shell closure, and
which cannot be properly described by a polynomial
expansion~\cite{[Dob93],[Sto07]}. In this kind of case, a good quality of the
Lipkin expansion obtained at small gauge angles is not sufficient enough to
guarantee a good convergence at larger gauge angles.
The situation is entirely different in the open-shell nucleus $^{120}$Sn, where
particle-number dependence of the projected energies is given by a smooth
function, which can very well be approximated by a polynomial expansion.
Here, for all gauge angles, we obtain a perfectly converging
Lipkin expansion of the exact energy kernel, even in the vicinity of the
pole related to the nearly half-filled 3s$_{1/2}$ orbital~\cite{[Dob07d]}.

\begin{figure}
\includegraphics[angle=0,width=0.9\columnwidth]{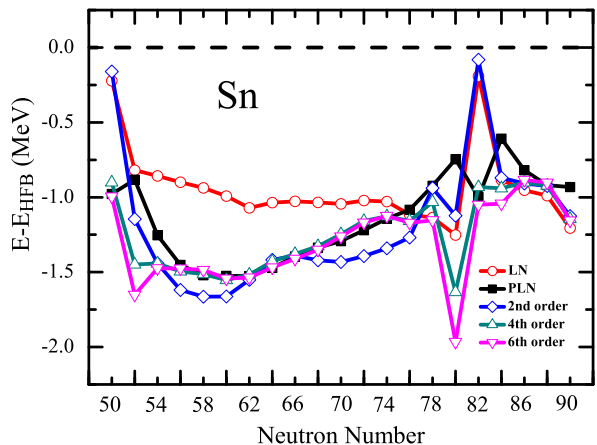}
\caption[T]{\label{fig:sn-lipkin}
(Color online) The LN, PLN, and Lipkin VAPNP energies of tin isotopes
relative to those obtained within the standard HFB method.
}
\end{figure}
\begin{figure}
\includegraphics[angle=0,width=0.9\columnwidth]{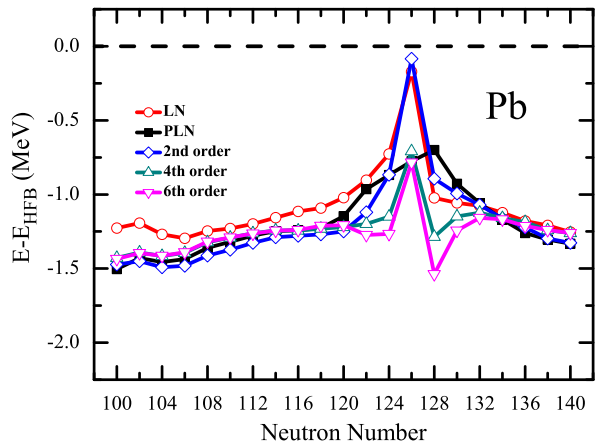}
\caption[T]{\label{fig:pb-lipkin}
(Color online) Same as in Fig.~\ref{fig:sn-lipkin}, but for lead isotopes.
}
\end{figure}

In Figs.~\ref{fig:sn-lipkin} and~\ref{fig:pb-lipkin}, we show results
of the Lipkin VAPNP method for tin and lead isotopes, respectively.
For comparison, figures also show results obtained
using the LN method, similarly as in Ref.~\cite{[Sto03]},
and PLN method, as in Ref.~\cite{[Sto07]}, where the exact PNP energy is
obtained via projection from HFB+LN self-consistent solution.

Away from the closed shells, at fourth and sixth orders, results of
the Lipkin VAPNP method are very similar to those of PLN results. As pointed
out by the VAPNP calculations in Ref.~\cite{[Sto07]}, for open shell
nuclei, PLN results are very close to exact VAPNP results. Again,
fourth and sixth orders give similar results, signaling the
convergence of the Lipkin expansion. We can thus conclude that the
fourth-order Lipkin VAPNP method is a good approximation of the exact
VAPNP method. Near shell closure, differences between
various orders of the Lipkin VAPNP method are large, indicating a
non-convergent power series of the Lipkin operator. Once again, this is
related to the kink in the particle-number dependence on the
projected energies~\cite{[Dob93],[Sto07]}.

\begin{figure}
\includegraphics[angle=0,width=0.9\columnwidth]{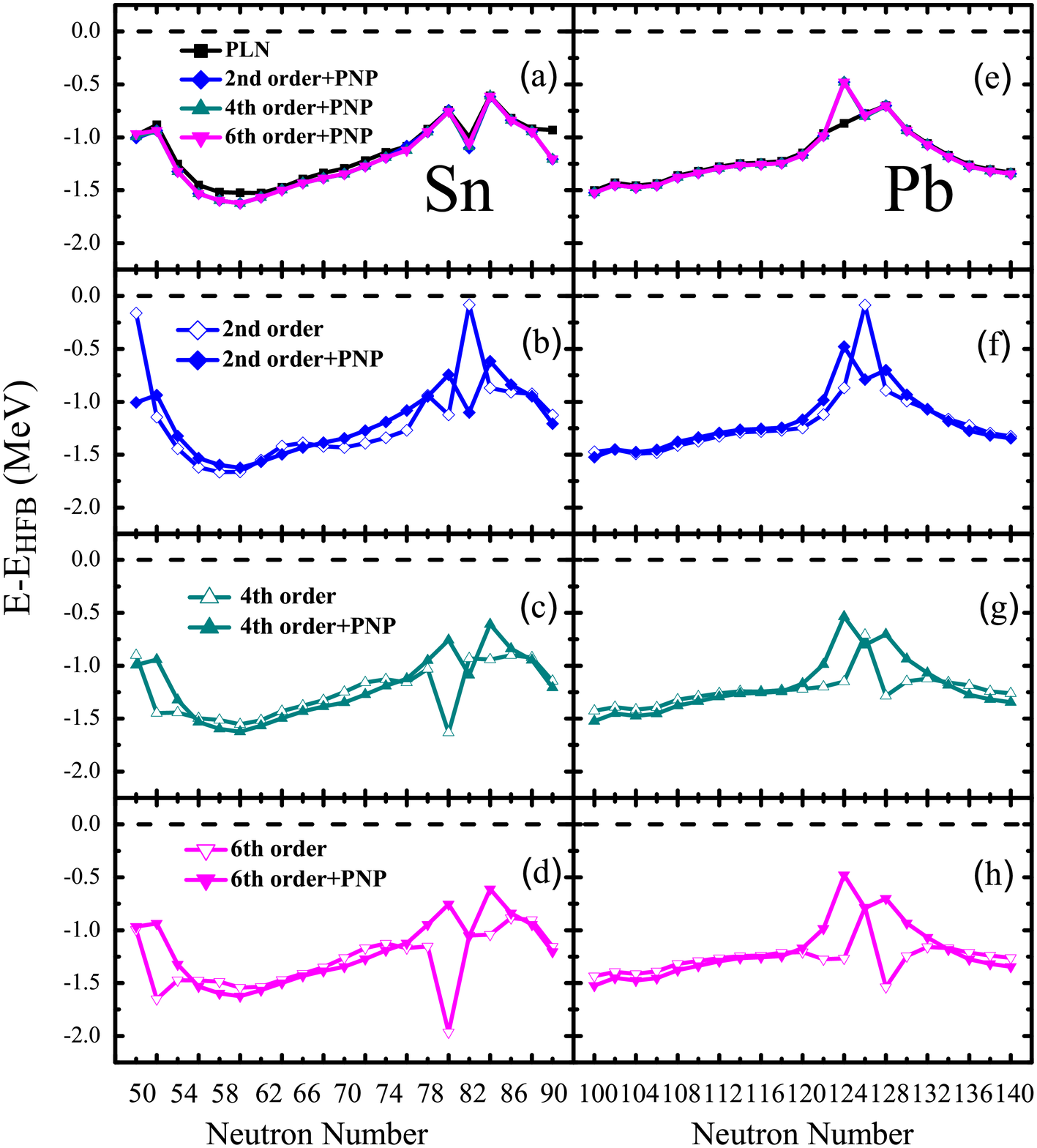}
\caption[T]{\label{fig:lipkinVAPNP}
(Color online)
Top panels (a) and (e): the PLN energies compared with the exact PNP
energies determined for states obtained by solving the Lipkin
equations at second, fourth, and sixth orders. Lower panels show
comparisons at three different orders of the Lipkin VAPNP  and exact
PNP energies. All energies are plotted relative to those obtained
within the standard HFB method.
}
\end{figure}

In Fig.~\ref{fig:lipkinVAPNP}, we show results obtained by projecting
good particle numbers from the states obtained either by the Lipkin VAPNP or LN
methods. It is very gratifying to see that irrespective on whether
one uses the Lipkin VAPNP or LN methods, the projected energies,
shown in the two top panels, are very similar. This fact means that
all approximate methods analyzed in this study lead to similar pair
condensates, whereas they differ in the determination of corrective
mean-field energies. The main advantage of using the Lipkin VAPNP
method is in the fact that the PNP calculation does not have to be
performed at all. Then, as shown in lower panels of
Fig.~\ref{fig:lipkinVAPNP}, the obtained energies very well
approximate the PNP energies. This is particularly true near the
middle of the shell, where the influence of closed-shell kinks in the
projected energies is weaker.

We also see that at closed shells, the second-order Lipkin VAPNP
method, similarly as the LN method -- see Figs.~\ref{fig:sn-lipkin}
and~\ref{fig:pb-lipkin} -- gives results that are very different from
those obtained by the PNP. On the contrary, the fourth- and
sixth-order Lipkin VAPNP methods gives results almost identical to
the PNP. Finally, at fourth and sixth orders, the non-analytic
behavior of the PNP energies at closed shells causes the largest
discrepancies for 2 or 4 particles away from the closed shell.

\begin{figure}
\includegraphics[angle=0,width=0.9\columnwidth]{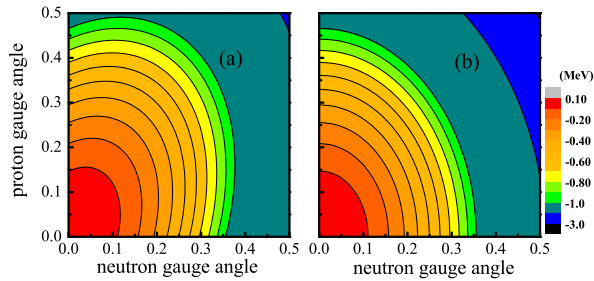}
\caption[T]{\label{fig:xe124}
(Color online)
Reduced energy kernel $h(\phi_{\nu},\phi_{\pi})-h(0,0)$ (left panel)
compared to the reduced kernel of the Lipkin operator
$\sum_{m=1}^M k_m(n_m(\phi_{\nu},\phi_{\pi})-n_m(0,0))$ at sixth order (right panel),
calculated for $^{124}$Xe.
}
\end{figure}

Our current implementation of the Lipkin method in the computer code {\sc HFODD}
allows us to treat pairing correlations simultaneously for neutrons
and protons. However, this has been implemented such that the Lipkin
operator is simply a sum of neutron and proton contributions
of Eq.~(\ref{eq:fN}), with Lipkin parameters determined by independent
gauge-angle rotations for neutrons and protons. Although this
implementation works perfectly well, we have realized that such a method
is insufficient in some cases. This is illustrated in Fig.~\ref{fig:xe124}(a),
which shows the reduced energy kernel of $^{124}$Xe calculated in two
dimensions, as a function of the neutron $\phi_n$ and proton $\phi_p$
gauge angles. We clearly see that energy kernel is tilted with
respect to main axes of neutron and proton gauge angle. Evidently, the
Lipkin operator, here being a sum of neutron and proton contributions separately,
leads to a non-tilted energy kernel, as shown in
Fig.~\ref{fig:xe124}(b). Therefore, to fully reproduce the true
energy kernel, one has to use the Lipkin operator that contains cross
terms, which depend on products of neutron and proton particle numbers.
Within proton-neutron pairing scheme, combined with PLN, these kind of
cross-terms are required~\cite{[Sat97w],[Sat00]}.
Implementations of such more complicated forms of the Lipkin operator
will be subject of future study.

\section{Summary}
\label{sec4}

In the framework of the nuclear energy-density-functional theory,
we derived the Lipkin method of approximate particle-number-symmetry
restoration up to sixth order. The Lipkin parameters were determined from
non-diagonal energy kernels, resulting in a more manageable approach
as compared to the traditional Lipkin-Nogami approach.

Convergence of the Lipkin VAPNP method was tested by investigating
gauge dependence of expansion parameters. Taking $^{120}$Sn as an example,
the Lipkin expansion up to the second order was found to have explicit gauge
angle dependence.
Inclusion of fourth order terms subsequently diminished dependence on the gauge angle
significantly. With the inclusion of sixth-order terms of the expansion, overall
change is minimal, indicating a converging series.
Accuracy of the Lipkin VAPNP method was tested by comparing the reduced energy
kernel and Lipkin operator approximated by a power series.
It was found that the chosen Lipkin operator describes well the small gauge-angle
rotation of the intrinsic wave function.
The results obtained for $^{100}$Sn and $^{120}$Sn showed that the second-order
Lipkin expansion is typically not sufficiently converged. Within fourth order,
the series already mimics the reduced energy kernels rather well.
With inclusion of sixth order term, the results stay practically the same,
indicating again a well converged series.

In chains of tin and lead isotopes, we compared the Lipkin VAPNP
method to LN and PLN methods. As pointed in Ref.~\cite{[Sto07]}, for
mid-shell nuclei, PLN is a very good approximation to the exact VAP
method. Our results show that for the mid-shell nuclei, the Lipkin
VAPNP method already at the second order gives rather well converged
results. When advancing to higher orders, the results are improved.
Near closed shells, because of the kink in the particle-number
dependence on the projected energy, the Lipkin VAPNP method is unable
to reproduce the exact projected energy. Also, near the closed-shell
region, the pairing correlations have dynamic nature~\cite{(Rin00)}.
Within the Lipkin VAPNP method, these kind of features cannot be
reproduced with a well converging series expansion. The main features
of the results obtained for nuclei were corroborated within the
exactly-solvable two-level pairing model.

When neutrons and protons were treated simultaneously within the
Lipkin VAPNP method, we observed a necessity to include in the Lipkin
operator cross terms, which depend simultaneously on neutron and
proton number operators. For the case of $^{124}$Xe, the contour
lines of the reduced energy kernel, with respect to neutron's and
proton's gauge angles, show tilted shapes. Without the cross terms,
this kind on behavior cannot be reproduced. A study of the cross
terms will be subject of future work.

The Lipkin VAPNP method presented in this work allows for a computationally
inexpensive way to approximate the exact VAPNP energy of the ground state. The
Lipkin method can be also applied to approximate the restoration of other
symmetries, broken on the mean-field level, with a small or no extra computational cost.
This is important, for example, in future adjustments of novel EDFs,
tailored for beyond mean-field multireference studies~\cite{[Dob12a]}.
Work towards restoring isospin and rotational symmetries within the framework
of the Lipkin method is currently in progress.

\begin{acknowledgments}
This work was supported in part by the Academy of Finland and
University of Jyv\"askyl\"a within the FIDIPRO programme, by the
Polish National Science Center under Contract No.~2012/07/B/ST2/03907,
by the Academy of Finland under the Centre of Excellence Programme
2012--2017 (Nuclear and Accelerator Based Physics Programme at JYFL),
by the European Union's Seventh Framework Programme ENSAR (THEXO)
under Grant No.~262010, and by the US Department of Energy under Contract Nos.\
DE-FC02-09ER41583 (UNEDF SciDAC Collaboration) and DE-FG02-96ER40963 (University
of Tennessee).
We acknowledge the CSC - IT Center for Science Ltd, Finland, for the
allocation of computational resources.
\end{acknowledgments}

\appendix

\section{Kernels and Lipkin parameters up to sixth order}
\label{app1}
To calculate kernels $N_m(\phi)$ from derivatives of the overlap
kernel $I(\phi)$, in Eq.~(\ref{eq:Nm}), we use the Onishi theorem~\cite{(Rin00)}:
\begin{eqnarray}
\langle\Phi|\Phi(\phi)\rangle &=&
\frac{\det^{1/2}(1+e^{2i\phi}CC^+)}
     {\det^{1/2}(1+          CC^+)}e^{-iN_0\phi} \, ,
\label{eq120a}
\end{eqnarray}
where $C$ is the Thouless matrix. Next, we use the identity
\begin{eqnarray}
\frac{d}{d\phi} \det(A) &=& \det(A) \Tr A^{-1} \frac{dA}{d\phi} \, ,
\label{eq121}
\end{eqnarray}
which is valid for any matrix $A$, provided that $\frac{dA}{d\phi}$
commutes with $A$, which is our case here.
This gives,
\begin{eqnarray}
\nonumber & &\frac{d}{d\phi} {\textstyle{\det^{1/2}}}(1+e^{2i\phi}CC^+)\\ &=&
 i {\textstyle{\det^{1/2}}}(1+e^{2i\phi}CC^+) \Tr \rho(\phi) \, ,
\label{eq122}
\end{eqnarray}
where
\begin{eqnarray}
\rho(\phi)  &=& e^{2i\phi}CC^+(1+e^{2i\phi}CC^+)^{-1} \, .
\label{eq123}
\end{eqnarray}
We then have the reduced kernels $n_m(\phi)=N_m(\phi)/I(\phi)$ given as
\begin{eqnarray}
n_1(\phi) &\!\!=\!\!&  \Tr \rho-N_0  \equiv   R_0 \, ,
\label{eq124} \\
n_2(\phi) &\!\!=\!\!&  R^2_0-i\frac{d}{d\phi}R_0   \equiv   R^2_0+R_1 \, ,
\label{eq125} \\
n_3(\phi) &\!\!=\!\!& R^3_0+3R_0R_1+R_2 \, ,
\label{eq126} \\
n_4(\phi) &\!\!=\!\!& R^4_0+6R^2_0R_1+4R_0R_2+3R^2_1+R_3 \, ,
\label{eq127}\\
%n_5(\phi) &\!\!=\!\!& R^5_0+10R^3_0R_1+10R^2_0R_2+15R_0R^2_1\nonumber\\
%               &&~~~~+5R_0R_3+10R_1R_2+R_4  \, ,
%\label{eq128}\\
%n_6(\phi) &\!\!=\!\!& R^6_0+15R^4_0R_1+20R^3_0R_2+45R^2_0R^2_1\nonumber\\
%               &&~~~~+15R^2_0R_3+60R_0R_1R_2+6R_0R_4\nonumber\\
%               &&~~~~+15R^3_1+15R_1R_3+10R^2_2+R_5 \, ,
n_5(\phi) & = & R^5_0+10R^3_0R_1+10R^2_0R_2+15R_0R^2_1\nonumber\\
          & &+5R_0R_3+10R_1R_2+R_4  \, ,
\label{eq128}\\
n_6(\phi) & = & R^6_0+15R^4_0R_1+20R^3_0R_2+45R^2_0R^2_1\nonumber\\
          & & +15R^2_0R_3+60R_0R_1R_2+6R_0R_4\nonumber\\
          & & +15R^3_1+15R_1R_3+10R^2_2+R_5 \, ,
\label{eq129}
\end{eqnarray}
where we used the definition
\begin{eqnarray}
R_n &=& (-i)^n\frac{d^n}{d\phi^n}\left(\Tr \rho - N_0\right) \, .
\label{eq130}
\end{eqnarray}
To lighten the notation, we omitted the arguments of $\rho(\phi)$ and $R_n(\phi)$
from here on.

Derivatives of the density matrix $\rho(\phi)$ (\ref{eq123}) can be calculated as,
\begin{eqnarray}
 -i\frac{d}{d\phi} \rho
          &=& 2 \rho(1-\rho) \, ,
\label{eq135} \\
 (-i)^2\frac{d^2}{d\phi^2} \rho
          &=& 4 \rho(1-\rho)(1-2\rho) \, ,
\label{eq136} \\
 (-i)^3\frac{d^3}{d\phi^3} \rho
          &=& 8 \rho(1-\rho) -48 \rho^2(1-\rho)^2 \, ,
\label{eq137}\\
 (-i)^4\frac{d^4}{d\phi^4} \rho
          &=& 16 \rho(1-\rho)^2 -16 \rho^2(1-\rho)\nonumber\\
          &-&192\rho^2(1-\rho)^3+192\rho^3(1-\rho)^2 \, ,
\label{eq138}\\
(-i)^5\frac{d^5}{d\phi^5} \rho
          &=& 32 \rho(1-\rho)^3 -128 \rho^2(1-\rho)^2\nonumber\\
          &+&32\rho^3(1-\rho)+2304\rho^3(1-\rho)^3\nonumber\\
          &-&768\rho^2(1-\rho)^4-768\rho^4(1-\rho)^2 \, ,
\label{eq139}
\end{eqnarray}
which gives
\begin{eqnarray}
 R_0      &=& \Tr \rho - N_0 \, ,
\label{eq144} \\
 R_1      &=& 2 \Tr\rho(1-\rho) \, ,
\label{eq145} \\
R_2       &=& 4 \Tr\rho(1-\rho)(1-2\rho) \, ,
\label{eq146} \\
R_3       &=& 8 \Tr\rho(1-\rho) -48\Tr \rho^2(1-\rho)^2 \, ,
\label{eq147} \\
R_4       &=& 16 \Tr\rho(1-\rho)^2-16 \Tr\rho^2(1-\rho)\nonumber\\
          &~&-192\Tr\rho^2(1-\rho)^3+192\Tr\rho^3(1-\rho)^2 \, ,
\label{eq148} \\
R_5       &=& 32 \Tr\rho(1-\rho)^3-128 \Tr\rho^2(1-\rho)^2\nonumber\\
          &~&+32 \Tr\rho^3(1-\rho)+2304\Tr\rho^3(1-\rho)^3\nonumber\\
          &~&-768\Tr\rho^2(1-\rho)^4-768\Tr\rho^4(1-\rho)^2 \, .
\label{eq149}
\end{eqnarray}

Up to the sixth order, the average Routhian of Eq. (\ref{eq:K-eig}) to be minimized reads
\begin{eqnarray}
E_{N_0}
 &=&\langle\Phi|\hat{H}-k_1(\hat{N}-N_0)-k_2(\hat{N}-N_0)^2
\nonumber \\\nonumber
 &&-k_3(\hat{N}-N_0)^3-k_4(\hat{N}-N_0)^4\\
 &&-k_5(\hat{N}-N_0)^5-k_6(\hat{N}-N_0)^6|\Phi\rangle \, .
\label{eq161}
\end{eqnarray}
Average values of powers of the particle-number operator are given in
Eqs.~(\ref{eq124})--(\ref{eq129}) taken at $\phi=0$. Moreover, in all terms
with $m\geq2$, one can set $R_0\equiv0$. This gives
\begin{eqnarray}
E_{N_0}
 &=&\langle\Phi|\hat{H}|\Phi\rangle-k_1(\Tr\rho-N_0)-2 k_2\Tr\rho(1-\rho)
\nonumber \\
 &&-4 k_3\Tr\rho(1-\rho)(1-2\rho)-12k_4(\Tr\rho(1-\rho))^2
\nonumber \\
 &&-8k_4\Tr\rho(1-\rho)+48 k_4 \Tr \rho^2(1-\rho)^2\nonumber \\
 &&-80k_5(\Tr\rho(1-\rho))(\Tr\rho(1-\rho)(1-2\rho))\nonumber \\
 &&-16k_5 \Tr \rho(1-\rho)^2+16k_5 \Tr \rho^2(1-\rho)\nonumber \\
 &&+192k_5\Tr \rho^2(1-\rho)^3-192k_5\Tr \rho^3(1-\rho)^2\nonumber \\
 &&-120k_6(\Tr\rho(1-\rho))^3-240k_6(\Tr\rho(1-\rho))^2\nonumber \\
 &&+1440k_6(\Tr\rho(1-\rho))(\Tr \rho^2(1-\rho)^2)\nonumber \\
 &&-160k_6(\Tr\rho(1-\rho)(1-2\rho))^2\nonumber \\
 &&-32k_6\Tr \rho(1-\rho)^3+128k_6\Tr \rho^2(1-\rho)^2\nonumber \\
 &&-32k_6\Tr \rho^3(1-\rho)-2304k_6\Tr \rho^3(1-\rho)^3\nonumber \\
 &&+768k_6\Tr \rho^2(1-\rho)^4+768k_6\Tr \rho^4(1-\rho)^2 \, .
\label{eq162}
\end{eqnarray}
Hence, the corresponding mean-field Routhians (to be used in the HFB equations)
reads
\footnote{Mean fields (\ref{eq163}) and reduced energy kernels
    (\ref{reduced-kernels}) are traditionally denoted by the same symbol $h$,
    but should not be confused one with another.}
\begin{eqnarray}
h'
 &=&h-k_1-\left(2 k_2+24k_4\Tr\rho(1-\rho)+8k_4)\right)(1-2\rho)
\nonumber \\
 &&-4 k_3\left((1-2\rho)^2-2\rho(1-\rho)\right)
\nonumber \\ &&
+96 k_4\rho(1-\rho)(1-2\rho) \nonumber \\ &&-80
k_5(\Tr\rho(1-\rho)(1-2\rho))(1-2\rho) \nonumber \\ &&-80
k_5(\Tr\rho(1-\rho))(1-2\rho)^2 \nonumber \\ &&
+160k_5(\Tr\rho(1-\rho))\rho(1-\rho)\nonumber \\ &&
-16k_5(1-\rho)^2+64k_5\rho(1-\rho)\nonumber \\ &&
-16k_5\rho^2+384k_5\rho(1-\rho)^3\nonumber \\
&&-1152k_5\rho^2(1-\rho)^2+384k_5\rho^3(1-\rho)\nonumber \\
&&-32k_6-360k_6(\Tr\rho(1-\rho))^2\nonumber \\
&&-320k_6(\Tr\rho(1-\rho)(1-2\rho))\nonumber \\
&&-480k_6(\Tr\rho(1-\rho))+1440k_6(\Tr\rho^2(1-\rho)^2)\nonumber \\
&&+1984k_6\rho+720k_6(\Tr\rho(1-\rho))^2\rho\nonumber \\
&&+1920k_6(\Tr\rho(1-\rho)(1-2\rho))\rho\nonumber \\
&&+3840k_6(\Tr\rho(1-\rho))\rho\nonumber \\
&&-2880k_6(\Tr\rho^2(1-\rho)^2)\rho-17280k_6\rho^2\nonumber \\
&&-8640k_6(\Tr\rho(1-\rho))\rho^2\nonumber \\
&&-1920k_6(\Tr\rho(1-\rho)(1-2\rho))\rho^2\nonumber \\
&&+49920k_6\rho^3+5760k_6(\Tr\rho(1-\rho))\rho^3\nonumber \\
&&-57600k_6\rho^4+23040k_6\rho^5 \, .
\label{eq163}
\end{eqnarray}

Lipkin parameters $k_{m}$ for $m=1,\ldots,M$ can be determined
from Eq.~(\ref{flat}) by requiring that it is fulfilled at gauge angle $\phi=\phi_0=0$
and also at all $M$ other nonzero values of the gauge angle $\phi_i$. This gives
\begin{eqnarray}
C+\sum_m k_m n_m(\phi_i)=h(\phi_i) \, ,
\label{eq164}
\end{eqnarray}
where $C$ is the flattened Routhian.
Then, at sixth order, Lipkin parameters $k_m$ can be easily obtained by inverting the matrix built
of coefficients $n_m(\phi_i)$ as
\begin{eqnarray}
&&\nonumber\left(\begin{array}{c}
C\\
k_1\\
k_2\\
k_3\\
k_4\\
k_5\\
k_6
\end{array}\right)=
\left(\begin{array}{cccc}
1&n_1(0)     &\cdots\,{}&n_6(0)     \\
1&n_1(\phi_1)&\cdots\,{}&n_6(\phi_1)\\
1&n_1(\phi_2)&\cdots\,{}&n_6(\phi_2)\\
1&n_1(\phi_3)&\cdots\,{}&n_6(\phi_3)\\
1&n_1(\phi_4)&\cdots\,{}&n_6(\phi_4)\\
1&n_1(\phi_5)&\cdots\,{}&n_6(\phi_5)\\
1&n_1(\phi_6)&\cdots\,{}&n_6(\phi_6)
\end{array}\right)^{-1}\left(\begin{array}{c}
h(0)\\
h(\phi_1)\\
h(\phi_2)\\
h(\phi_3)\\
h(\phi_4)\\
h(\phi_5)\\
h(\phi_6)
\end{array}\right) \, .\\
\label{eq165}
\end{eqnarray}
At lower orders, or when neglecting odd orders,
a smaller number of the gauge angle points can be used.

In fact, the value of $C$ obtained from the first row of
Eq.~(\ref{eq165}) does not appear in the mean-field Routhian
(\ref{eq163}) and can be ignored. Anyhow, at convergence it is
calculated from Eq.~(\ref{eq162}). Moreover, the value of $k_{1}$
obtained from the second row of Eq.~(\ref{eq165}) can be ignored too,
because, when iterating the HFB equation, it is anyhow used to properly
adjust the average particle number, and, at convergence, one has
$\Tr\rho=N_0$ and thus it has no influence on the value of the right-hand side of
Eq.~(\ref{eq162}).

\section{Lipkin method applied to the two-level pairing model}
\label{app2}

In this appendix, we apply the Lipkin method to the standard two-level
pairing model, which is characterized by two $\Omega$-fold degenerate
levels with the single-particle energy difference $2\epsilon$ and
pairing strength $G$. Below we closely follow the notations
and definitions presented in Refs.~\cite{[Zhe92],[Dob93]}, where
the results obtained within the LN method have been studied.

\begin{figure}[h]
\includegraphics[angle=0,width=0.9\columnwidth]{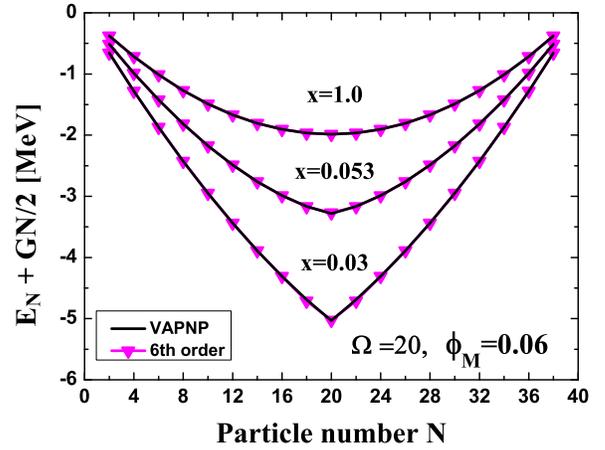}
\caption[T]{\label{fig:E_N}
(Color online) Ground-state energies in the two-level pairing model
calculated within the sixth-order Lipkin and exact VAPNP methods. To render the
curves symmetric with respect to the closed shell at $N=20$, the
appropriate linear term was added. Normalization of
$50G+\epsilon=1$\,MeV was used.
}
\end{figure}

\begin{figure}
\includegraphics[angle=0,width=0.9\columnwidth]{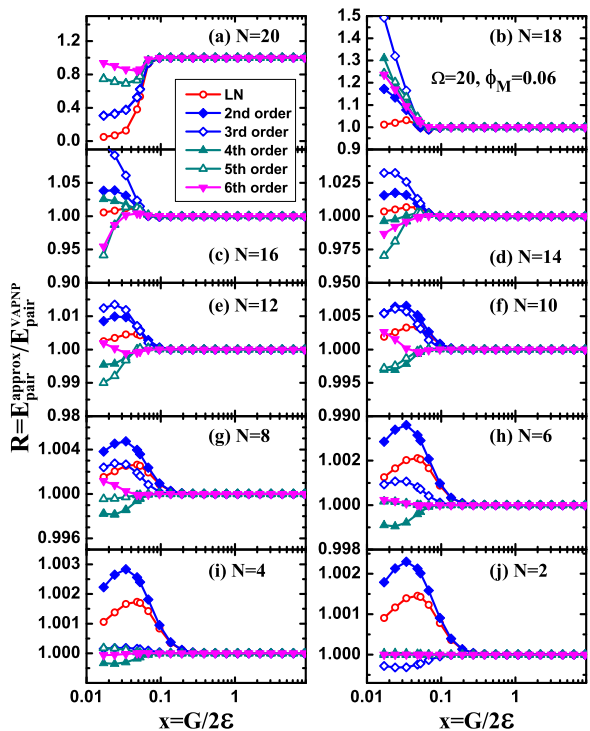}
\caption[T]{\label{fig:R&G}
(Color online) Ratios of approximate pairing energies,
calculated within the approximate LN (open circles) and
Lipkin VAPNP methods, relative to those of the exact VAPNP method. The figure shows results
obtained for $\phi_{M}=0.06$ in function of the
pairing-strength parameter $x=G/2\epsilon$. Note that
panels (a)--(j) are drawn in very different scales, indicating the
discrepancies up to 100\% for $N=20$ and only 0.2\% for
$N=2$.
}
\end{figure}

In Fig.~\ref{fig:E_N}, we show particle-number dependence of the
ground-state energies obtained for $\Omega=20$ and for three values
of the ratio $x=G/2\epsilon$ equal to 0.03 (weak pairing) 0.053
(critical pairing), and 1 (strong pairing). Results show excellent
agreement between the sixth-order Lipkin and exact VAPNP methods,
which in the absolute scale of energy cannot be distinguished one
from another. To compare the approximate and exact VAPNP methods in
fine detail, in Fig.~\ref{fig:R&G} we plotted ratios of the respective
pairing energies,
$R=E^{\text{approx}}_{\text{pair}}$/$E^{\text{VAPNP}}_{\text{pair}}$,
as functions of $x$. The pairing energies are defined~\cite{[Zhe92],[Dob93]}
as differences between the total and Hartree-Fock energies.
Note that the results are exactly symmetric with
respect to the mid shell, that is, those for particle numbers of $N$
and $2\Omega-N$ are exactly identical.

The ratio of $R=1$, that is, perfect agreement, is for all particle
numbers reached in the strong-pairing regime. For weak pairing, the
largest discrepancies appear at mid shell, $N=20$, and they gradually
decrease towards smaller (or larger) particle numbers. This is
related to the kink in particle-number dependence of ground-state
energies~\cite{[Dob93]}, cf.\ Fig.~\ref{fig:E_N}, which disappears
with increasing pairing correlations.

For $N=20$, with increasing order of the Lipkin expansion, the
agreement with exact results gradually increases, and the Lipkin
VAPNP method, even at second order, is here visibly superior to the
LN method. Note that at $N=20$, the odd orders of expansion (third and fifth)
do not bring any improvement -- this is owing to the symmetry of the model
with respect to the mid shell.

For $N=18$, the Lipkin expansion cannot reproduce the kink appearing
at the adjacent particle number of $N=20$ (see Fig.~\ref{fig:E_N}), and it does not seem to
converge to the exact result, whereas the LN results are clearly
superior. For smaller particle numbers, this pattern gradually changes,
and for $N\leq12$ the Lipkin expansion does converge to the exact
result and at orders higher than four becomes better than the LN method.

We stress here that in the realistic cases discussed in Sec.~\ref{sec3},
the pattern of comparison between the LN and Lipkin VAPNP methods
pertains to moderately high pairing strengths, certainly beyond
the pairing phase transition, which in the two-level model appears at
$x_c=1/(\Omega-1)\simeq0.053$.

\begin{figure}
\includegraphics[angle=0,width=0.92\columnwidth]{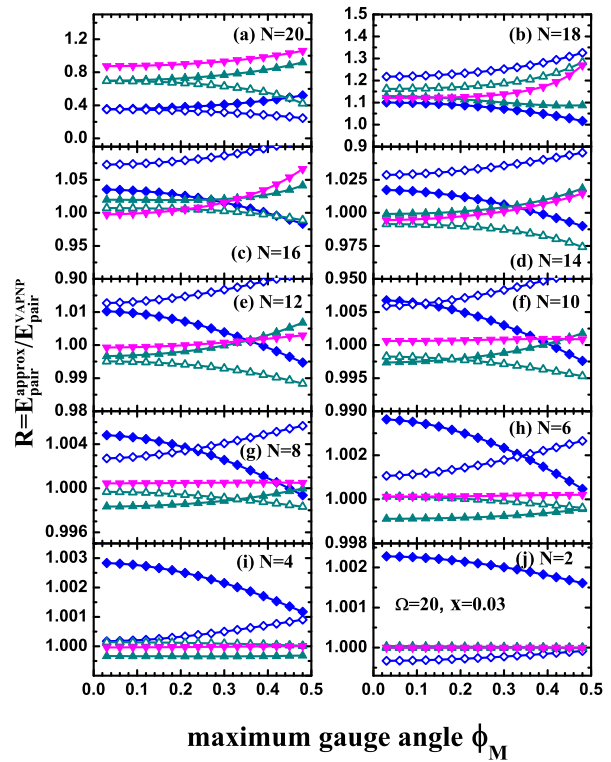}
\caption[T]{\label{fig:R&phi_max}
(Color online) Same as in Fig.~\protect\ref{fig:R&G}, but for the
results obtained for $x=0.03$ (weak pairing) plotted in function of the
maximum gauge angle $\phi_{M}$.
}
\end{figure}

Finally, in Fig.~\ref{fig:R&phi_max}, we show dependence of results
on the maximum gauge angle $\phi_{M}$ used in the Lipkin VAPNP
method, see Secs.~\ref{sec2} and~\ref{sec3}. We see that for all
particle numbers, the second-order results do depend on $\phi_{M}$,
indicating an insufficient order of expansion. For $N\leq12$ we see
that with increasing order of expansion, the results become perfectly
independent of $\phi_{M}$, which characterizes a converging
expansion. On the other hand, closer to the mid shell, even at sixth
order a visible dependence on $\phi_{M}$ still remains.

\bibliographystyle{unsrt}
%\bibliography{lipkin-pnp,jacwit31}
%\bibliography{lipkin-pnp,C:/Actual/LaTeX/Latex.all/jacwit31}

\end{document}